\documentstyle[12pt]{article}
\def\nabstar#1{\nabla\kern-0.5pt\smash{\raise 4.5pt\hbox{$\ast$}}
               \kern-4.5pt_{#1}}

\def\drvstar#1{\partial\kern-0.5pt\smash{\raise 4.5pt\hbox{$\ast$}}
               \kern-5.0pt_{#1}}

\def\newline{\relax\ifhmode\null\hfil\break\else\nonhmodeerr@\newline\fi}
\def\frac#1#2{{#1\over#2}}
\def\text#1{{\hbox{\rm #1}}}

\newcommand{\beq}{\begin{equation}}
\newcommand{\eeq}{\end{equation}}
\newcommand{\bea}{\begin{eqnarray}}
\newcommand{\eea}{\end{eqnarray}}
\def\Id{ \mbox{1\hspace{-1.2mm}I} }
\def\CC{ \mbox{C\hspace{-2.8mm}C} }
\def\BE{\begin{equation}}
\def\EE{\end{equation}}
\def\BA{\begin{eqnarray}}
\def\EA{\end{eqnarray}}
\def\BAN{\begin{eqnarray*}}
\def\EAN{\end{eqnarray*}}

\def\nn{\nonumber\\}

\def\gm5{\gamma_5}

\input epsf.sty
\newdimen\psfigsize
\def\psfigure#1 #2 #3 #4 #5{
    \begin{figure}[tbh]
      \begin{center}
      \vbox{
        \null\vskip-0.2in\hskip#2
        \epsfxsize=#1
        \epsfbox{#4}
        \vskip -0.3in
        \caption {#5 \label{#3}}
        \vskip 0.0 true in plus 0.3 true in
      }
      \end{center}
   \end{figure}
}
\begin{document}
\thispagestyle{empty}
\begin{flushright}
NTUTH-02-505C \\
June 2002 \\
\end{flushright}
\bigskip\bigskip\bigskip
\vskip 2.5truecm
\begin{center}
{\LARGE {A note on the Zolotarev optimal rational approximation
for the overlap Dirac operator}}
\end{center}
\vskip 1.0truecm
\centerline{Ting-Wai Chiu, Tung-Han Hsieh, Chao-Hsi Huang, Tsung-Ren Huang}
\vskip5mm
\centerline{Department of Physics, National Taiwan University}
\centerline{Taipei, Taiwan 106, Taiwan.}
\centerline{\it E-mail : twchiu@phys.ntu.edu.tw}
\vskip 1cm
\bigskip \nopagebreak \begin{abstract}
\noindent

We discuss the salient features of Zolotarev optimal rational
approximation for the inverse square root function, in particular,
for its applications in lattice QCD with overlap Dirac quark.
The theoretical error bound for the matrix-vector
multiplication $ H_w (H_w^2)^{-1/2}Y $ is derived.
We check that the error bound is always satisfied
amply, for any QCD gauge configurations we have tested.
An empirical formula for the error bound is determined,
together with its numerical values
( by evaluating elliptic functions )
listed in Table \ref{table:drZ} as well as
plotted in Figure \ref{fig:dev_zolo}.
Our results suggest that with Zolotarev approximation
to $ (H_w^2)^{-1/2} $, one can practically preserve the
exact chiral symmetry of the overlap Dirac operator
to very high precision, for any gauge configurations
on a finite lattice.



\end{abstract}
\vskip 1.5cm

\newpage\setcounter{page}1

\section{Introduction}

In lattice QCD with overlap Dirac quarks
\cite{Neuberger:1998fp,Narayanan:1995gw},
one encounters the challenging problem of taking the inverse square
root of a positive definite Hermitian matrix, which stems from the
overlap Dirac operator for massless fermion
\bea
\label{eq:Dh}
D = m_0 a^{-1} \left( 1 + \gamma_5 \frac{H_w}{\sqrt{H_w^2}} \right) \ ,
\eea
where $ H_w $ denotes the Hermitian Wilson-Dirac operator with a
negative parameter $ -m_0 $,
\bea
H_w = \gamma_5 D_w = \gamma_5 ( -m_0 + \gamma_\mu t_\mu + W ) \ ,
\eea
$ \gamma_\mu t_\mu $ the naive fermion operator, and $ W $ the Wilson term.
It is well-known that (\ref{eq:Dh}) is the simplest and the most
elegant realization of chiral symmetry
on a finite lattice \cite{Ginsparg:1982bj}
\bea
\label{eq:gwr}
D\gamma_5+\gamma_5 D=a m_0^{-1} D \gamma_5 D \ ,
\eea
and it also fulfils all basic requirements
( i.e., doublers-free, $ \gamma_5$-hermiticity, exponentially-local,
and correct continuum behavior ) for a decent lattice Dirac operator.
However, in practice, how to compute physical quantities in lattice
QCD with overlap Dirac operator (\ref{eq:Dh}) is still a challenging
problem, due to the inverse square root of $ H_w^2 $ in (\ref{eq:Dh}),
which cannot be evaluated analytically in a closed form, nor numerically
via diagonalization since the memory requirement exceeds the physical
memory of the present generation of computers. Thus, at this stage,
it is necessary to replace $ (H_w^2)^{-1/2} $ with a good approximation,
in any lattice QCD calculations with overlap Dirac quark.
The relevant question is what is the optimal rational approximation
$ r_n (x) $ ( of degree $ n $ ) for the
inverse square root function $ x^{-1/2}, x \in [1,b] $, in view of the
convergence of a rational approximation can attain
\bea
\label{eq:dev}
\max_{ 1 \le x \le b } | 1 - \sqrt{x} r_{n}(x) | =
 c_1 e^{- c_2 n } \ ,
\hspace{6mm}   c_1, \ c_2 > 0 \ ,
\eea
which is much faster than that of any polynomial approximation.
It turns out that the optimal solution has been obtained by
Zolotarev \cite{Zolotarev:1877} more than 100 years ago.
A detailed discussion of Zolotarev's result can be found in
Akhiezer's two books \cite{Akhiezer:1992,Akhiezer:1990}.
Unfortunately, Zolotarev's optimal rational
approximation has been overlooked by the numerical algebra
community until recent years\footnote{
See, for example, Ref. \cite{Ingerman:2000ab}.}.

A comparative study of Zolotarev's approximation versus other schemes
for computing the product of the matrix sign function $ H_w ( H_w^2 )^{-1/2} $
with a vector $ Y $, has been reported in Ref. \cite{vandenEshof:2002ms}.

Recently, we have used Zolotarev optimal rational approximation to
compute overlap Dirac quark propagator in the quenched approximation
\cite{Chiu:2002xm}. The parameters of the pseudoscalar
meson mass formula to one-loop order in chiral perturbation theory are
determined, and from which the light quark masses $ m_{u,d} $ and $ m_s $
can be extracted \cite{Chiu:2002rk} with the experimental inputs
of pion and kaon masses, and the pion decay constant.

Nevertheless, in Refs. \cite{vandenEshof:2002ms,Chiu:2002xm},
the underlying principles and salient features\footnote{
In particular, the theoretical error bound for the product
$ H_w ( H_w^2 )^{-1/2} Y $ can be derived in terms of the
deviation (\ref{eq:drZ}), as shown in Section 5.
Moreover, we clarify that the coefficients
$ d_0 $ (\ref{eq:d0}) and $ D_0 $ (\ref{eq:D0})
can be obtained explicitly, {\it without} using the
condition : $ \max [ 1 - \sqrt{x} r_{n}(x) ]|_{ 1 \le x \le b } =
- \min [ 1 - \sqrt{x} r_{n}(x) ]|_{ 1 \le x \le b } $.}
of Zolotarev optimal rational approximation have not been addressed.
Also, in Refs. \cite{Ingerman:2000ab, vandenEshof:2002ms, Chiu:2002xm},
only one option (\ref{eq:rZ'}) of the two possible choices [(\ref{eq:rZ})
and (\ref{eq:rZ'})] of Zolotarev's rational polynomials has been exploited,
however, these two options are complementary to each other,
and in some cases it is more effective to use (\ref{eq:rZ}) than
(\ref{eq:rZ'}), in computing $ (H_w^2)^{-1/2} $ times a column vector $ Y $,
as discussed in Section 5.

In this paper, we discuss the salient features of Zolotarev optimal
rational approximation, in particular, for its application in lattice
QCD with overlap Dirac operator. We start with the basic principles of
rational approximation, and show that Zolotarev's rational polynomial
is indeed the optimal rational approximation for the inverse square root
function. Then we examine how well the Zolotarev optimal rational
approximation (\ref{eq:zolo}) for the overlap Dirac operator can preserve
the exact chiral symmetry (\ref{eq:gwr}) on a finite lattice.
For the overlap Dirac operator, (\ref{eq:gwr}) is equivalent to
\bea
\label{eq:gwr_Z}
\left( \frac{H_w}{\sqrt{H_w^2}} \right)^2 = 1 \ .
\eea
Thus the question is how much the exact relation (\ref{eq:gwr_Z})
is violated if one replaces $ (H_w^2)^{-1/2} $ with Zolotarev's
optimal rational polynomial $ r^{(n)}_Z(H_w^2) $.
The chiral symmetry breaking can be measured in terms of
the deviation
\bea
\label{eq:Delta_Z}
\Delta_Z = \max_{ \forall \ Y \ne 0 }
\left|\frac{Y^{\dagger}\{H_w r^{(n)}_Z(H_w^2)\}^2 Y}{Y^{\dagger}Y}-1 \right| \ .
\eea

It turns out that the deviation (\ref{eq:Delta_Z}) has a theoretical upper
bound (\ref{eq:DZB}), which is a function of $ n $ ( the degree
of the rational polynomial ), and $ b = \lambda_{max}^2 / \lambda_{min}^2 $,
where $ \lambda_{max}^2 $ and $ \lambda_{min}^2 $ denote the maximum
and the minimum of the eigenvalues of $ H_w^2 $.
Thus, for any given gauge configuration, one can use
the theoretical upper bound to determine what values
of $ n $ and $ b $ ( i.e., how many low-lying eigenmodes of $ H_w^2 $
should be projected out ) are required to attain one's
desired accuracy in preserving the chiral symmetry
of the overlap Dirac operator.
In practice, one has no difficulties to achieve
$ \Delta_Z < 10^{-12} $ for any gauge configurations
on a finite lattice.

The outline of this paper is as follows. In Sections 2-4,
we review the basic principles of rational approximation,
and show that Zolotarev's rational polynomial is indeed the optimal
rational approximation for the inverse square root function.
In Section 5, we discuss the Zolotarev approximation
for $ (H_w^2)^{-1/2} $ in the overlap Dirac operator,
and derive the theoretical error bound for the matrix-vector
multiplication $ H_w (H_w^2)^{-1/2} Y $ ( $ Y $ : any nonzero column vector ),
which is the most essential operation in computing the propagator of
overlap Dirac quark by nested conjugate gradient. We check that the
theoretical error bound is always satisfied amply, for any QCD gauge
configurations we have tested. Then the numerical values of
the theoretical error bound are computed, and listed in Table \ref{table:drZ},
as well as plotted in Figure \ref{fig:dev_zolo}.
An empirical formula for the error bound is
determined from the numerical data.
In Section 6, we conclude with some remarks.

\section{de la Vall\'{e}e-Poussin's theorem}

First we consider the following problem.
Given any two positive and continuous functions $ f(x) $ and $ g(x) $
for $ x \in [1, b] $, the problem is to find the irreducible
rational polynomial of degree $ n $ with positive real coefficients,
\bea
\label{eq:rn}
r(x) = \frac{p(x)}{q(x)}
     = \frac{ p_n x^n + p_{n-1} x^{n-1} + \cdots + p_0 }
            { q_n x^n + q_{n-1} x^{n-1} + \cdots + q_0 } \ ,
\eea
such that the deviation of $ g(x) r(x) $ from $ f(x) $ is the
minimum. Here the deviation is defined as the
maximum of $ | f(x) - g(x) r(x) | $ for the
entire interval $ [ 1, b ] $, i.e.,
\bea
d(r) = \max_{ x \in [1,b] } | f(x) - g(x) r (x) | \ .
\eea

Now if there exists $ r(x) $ such that the difference
\bea
\delta(x) = f(x) - g(x) r(x)
\eea
changes sign alternatively $ 2 n + 2 $ times within
the interval $ [ 1 , b ] $, and attains its maxima and minima, say,
\bea
\delta(x) =  \delta_1, -\delta_2, \cdots, +\delta_{2n+1},
            -\delta_{2n+2} \hspace{6mm} ( \delta_i > 0 )
\eea
at consecutive points,
\bea
x_1 < x_2 < \cdots < x_{2n+2} \ ,
\eea
then it can be shown that
\bea
\label{eq:dh}
d(r) \ge \min \{ \delta_1, \delta_2, \cdots, \delta_{2n+2} \} \equiv d_{min}
\eea
for all irreducible rational polynomial $ r(x) $ as defined by (\ref{eq:rn}).
This can be asserted as follows.

The strategy is to assume the contrary is true, and then show that it
leads to contradiction. Suppose there exists an irreducible rational
polynomial $ R(x) $ of degree $ n $, such that $ d(R) < d_{min} $.
Then
\bea
\label{eq:D}
D(x) &\equiv & g(x) R(x) - g(x) r(x)    \nn
     & = & f(x) - g(x) r(x) - [ f(x) - g(x) R(x) ]  \nn
     & = & \delta(x) - \Delta(x)
\eea
is a continuous function, and it also changes sign alternatively
at least $ 2 n + 2 $ times in the interval $ [1,b] $, since
$ | \Delta(x) | \le d(R) < d_{min} $.
Thus $ D(x) $ has at least $ 2 n + 1 $ zeros in the interval $ (1, b) $.
On the other hand,
\bea
D(x) = g(x) [ R(x) - r(x) ] \equiv g(x) \frac{u(x)}{v(x)}
\eea
where $ u(x) $ and $ v(x) $ are polynomials of degree $ 2n $.
Therefore $ D(x) $ cannot have more than $ 2n $ zeros,
i.e., a contradiction.
This completes the proof of de la Vall\'{e}e-Poussin's theorem.

In general, de la Vall\'{e}e-Poussin's theorem\cite{Akhiezer:1992}
asserts that if there exists an irreducible rational polynomial of
the form
\bea
\label{eq:rnm}
r^{(n,m)}(x)=
\frac{ p_{n} x^{n} + p_{n-1} x^{n-1} + \cdots + p_0 }
     { q_{m} x^{m} + q_{m-1} x^{m-1} + \cdots + q_0 } \ ,
     \ ( m \ge n, \ p_i, q_i > 0 )
\eea
such that $ \delta(x) = f(x) - g(x) r^{(n,m)}(x) $ has
$ n + m + 2 $ alternate change of sign in the interval $ [ 1, b ] $,
and attains its maxima and minima, say,
\BAN
\delta(x) =  \delta_1, -\delta_2, \cdots,
            (-1)^{n+m+1}\delta_{n+m+2} \hspace{6mm} ( \delta_i > 0 )
\EAN
at consecutive points,
\BAN
x_1 < x_2 < \cdots < x_{n+m+2} \ ,
\EAN
then
\BAN
d(r) \ge \min \{ \delta_1, \delta_2, \cdots, \delta_{n+m+2} \} \equiv d_{min}
\EAN
for all irreducible rational polynomial $ r^{(n,m)}(x) $.

In the next section, we show that there exists an optimal $ r_Z(x) $ such
that $ d(r_Z) $ is minimum, i.e., Zolotarev's solution \cite{Zolotarev:1877}.

\section{Zolotarev's optimal rational approximation}

First, we recall some well-known properties of Jacobian elliptic functions
\cite{Akhiezer:1990}. The Jacobian elliptic function
$ \mbox{sn}(u;\kappa) = \eta $ is defined by the elliptic integral
\bea
\label{eq:sn}
u(\eta) = \int_{0}^{\eta} \frac{dt}{\sqrt{(1-t^2)(1- \kappa^2 t^2 )}} \ .
\eea
It is a meromorphic function of $ u $ and $ \kappa $, which is defined
over $ \CC^2 $.
For $ \eta = 1 $, (\ref{eq:sn}) becomes
\bea
u(1) \equiv K = \int_{0}^{1} \frac{dt}{\sqrt{(1-t^2)(1- \kappa^2 t^2 )}} \ ,
\eea
the complete elliptic integral of the first kind with modulus $ \kappa $.

The important feature of $ \mbox{sn}(u;\kappa) $ is that it is periodic
in both real and imaginary parts of $ u $,
\bea
\mbox{sn}(u + 4K ;\kappa) = \mbox{sn}(u;\kappa) \\
\mbox{sn}(u + 2 i K' ;\kappa) = \mbox{sn}(u;\kappa)
\eea
where $ K' $ is complete elliptic integral of the first kind
with modulus $ \kappa' = \sqrt{ 1 - \kappa^2 } $

Now if we define
\bea
\label{eq:x}
x(u;\kappa) \equiv \mbox{sn}^2 (u;\kappa) \ ,
\eea
then $ x $ has periods $ 2 K $ and $ 2 i K' $, since
$ \mbox{sn}(u + 2K ;\kappa) = - \mbox{sn}(u;\kappa) $.

The crucial formula for Zolotarev's optimal rational approximation
is the second principal $n$-th degree transformation of
Jacobian elliptic function \cite{Akhiezer:1990}.
In terms of $ x $ (\ref{eq:x}), it reads
\bea
\label{eq:transf}
\sqrt{ x \left( \frac{u}{M};\lambda \right) } =
\sqrt{ x (u;\kappa) } \frac{1}{M}
\prod_{l=1}^{n} \frac{ 1 + x (u;\kappa)/c_{2l} }
                     { 1 + x (u;\kappa)/c_{2l-1} }
\eea
where
\bea
\label{eq:lambda}
\lambda &=& \prod_{l=1}^{2n+1}
\frac{\Theta^2 \left(\frac{2lK'}{2n+1};\kappa' \right)}
     {\Theta^2 \left(\frac{(2l-1)K'}{2n+1};\kappa' \right)} \ , \\
\label{eq:M}
M &=& \prod_{l=1}^{n}
\frac{\mbox{sn}^2 \left(\frac{(2l-1)K'}{2n+1};\kappa' \right)}
     {\mbox{sn}^2 \left(\frac{2lK'}{2n+1};\kappa' \right)} \ , \\
\label{eq:cl}
c_l &=& \frac{\mbox{sn}^2(\frac{lK'}{2n+1}; \kappa' ) }
             {1-\mbox{sn}^2(\frac{lK'}{2n+1}; \kappa' )},
\eea
and $ \Theta $ denotes the elliptic theta function.
Note that $ x( u/M ; \lambda ) $ has periods $ 2 L = 2 K/M $ and
$ 2 i L' = 2 i K'/ [ ( 2 n + 1 ) M ] $,
\bea
x \left(\frac{u + 2K}{M};\lambda \right) &=&
x \left( \frac{u}{M};\lambda \right) \\
x \left(\frac{u}{M} + \frac{2iK'}{(2n+1)M} ;\lambda \right) &=&
x \left( \frac{u}{M};\lambda \right) \ .
\eea

Now restricting $ u = K + i v $, we have
\bea
\label{eq:xz}
x(u;\kappa) = \mbox{sn}^2 ( u ; \kappa )
= \frac{ 1 - \mbox{sn}^2 ( iv; \kappa )}
       { 1 -  \kappa^2 \mbox{sn}^2 ( iv; \kappa) }
= \frac{1}{ 1 -  {\kappa'}^2 \mbox{sn}^2 ( v; \kappa') } \ ,
\eea
where the identity
\bea
\mbox{sn}^2 ( iv; \kappa ) = - \frac{ \mbox{sn}^2 ( v; \kappa' ) }
                             { 1 - \mbox{sn}^2 ( v; \kappa' ) }
\eea
has been used.
Equation (\ref{eq:xz}) implies that $ x $ increases from
$ 1 $ to $ 1/\kappa^2 $ as $ v $ increases from $ 0 $ to $ K' $,
since $ \mbox{sn} ( 0 ; \kappa') = 0 $, and
$ \mbox{sn} ( K' ; \kappa') = 1 $, from the definition (\ref{eq:sn}).

Now consider
\bea
\label{eq:rZ}
r_Z (x) = \frac{2 \lambda}{ 1 + \lambda } \frac{1}{M}
\prod_{l=1}^{n} \frac{ 1 + x/c_{2l} }{ 1 + x/c_{2l-1} }
\eea
as a rational approximation to $ 1/\sqrt{x} $.
Then the deviation can be obtained from
\bea
\label{eq:delta_Z}
\delta_Z(x) = 1 - \sqrt{x} r_Z (x)
            = 1 - \frac{ 2 \lambda }{ 1 + \lambda }
                   \mbox{sn} \left( \frac{K+iv}{M};\lambda \right)
\eea
where (\ref{eq:transf}) has been used.
Using identity (\ref{eq:xz}) with substitutions
$ \kappa \leftarrow \lambda $ and
$ u \leftarrow (K+iv)/M $,
then (\ref{eq:delta_Z}) becomes
\bea
\delta_Z(x) = 1 - \frac{ 2 \lambda }{1 + \lambda} \frac{1}
{\sqrt{1-{\lambda'}^2 \mbox{sn}^2 \left( \frac{v}{M};\lambda' \right)}} \ ,
\eea
where $ \lambda' = \sqrt{ 1 - \lambda^2 } $.
As $ v $ changes from $ 0 $ to $ K' $, $ v/M $ changes from $ 0 $
to $ K'/ M = ( 2 n + 1 ) L' $. This implies that
$ \mbox{sn}^2 \left( v/M ; \lambda' \right) $
attains its minima and maxima alternatively as $ 0, 1, \cdots, 0, 1 $ at
$ v/M = 0, L', \cdots, 2n L', (2n+1)L' $.
Correspondingly,
$ \delta_Z(x) $ reaches its maxima and minima alternatively as
\bea
\label{eq:max_min}
\frac{1-\lambda}{1+\lambda}, -\frac{1-\lambda}{1+\lambda}, \cdots,
\frac{1-\lambda}{1+\lambda}, -\frac{1-\lambda}{1+\lambda}
\eea
at $ v/M = 0, L', \cdots, 2n L', (2n+1)L' $, which
correspond to $ 2n + 2 $ successive $ x \in [ 1, 1/\kappa^2 ] $,
\bea
1,  \frac{1}
        {1-\kappa'^2 \mbox{sn}^2 \left(\frac{K'}{2n+1}; \kappa' \right)},
\cdots,
\frac{1}
{ 1- \kappa'^2 \mbox{sn}^2 \left( \frac{2n K'}{2n+1}; \kappa' \right)},
\frac{1}{\kappa^2} \ .
\eea
That is,
\bea
\label{eq:xi}
x_i = \frac{1}
{1-\kappa'^2 \mbox{sn}^2 \left(\frac{(i-1) K'}{2n+1}; \kappa' \right)} \ , \
i=1, \cdots, 2n + 2 \ .
\eea
From (\ref{eq:max_min}), we conclude that
\bea
\label{eq:drZ}
d(r_Z) = \frac{1-\lambda}{1+\lambda} = d_{min} \ ,
\eea
and Zolotarev's solution (\ref{eq:rZ}) is the optimal rational
approximation for the inverse square root function, according to
de la Vall\'{e}e-Poussin's theorem.

At this point, it is instructive to plot
$ \delta_Z(x) = 1 - \sqrt{x} r_Z(x) $ explicitly, and to
see how it attains its maxima and minima alternatively.
This is shown in Fig. \ref{fig:zolo_n6} for
$ n = 6 $ and $ b = \kappa^{-2} = 1000 $.
Note that there are exactly $ 2n + 2 = 14 $ alternate change
of sign in $ [ 1, 1000 ] $, with maxima and minima at
\BAN
x &=& 1, \ 1.145, \ 1.664, \ 2.858, \ 5.415, \ 10.80, \ 22.05, \ 45.34, \\
  & & 92.59, \ 184.7, \ 349.9, \ 600.9, \ 873.3, \ 1000.
\EAN

In Fig. \ref{fig:xz}, we plot the positions $ x_i $ (\ref{eq:xi}) of the
maxima and minima of $ \delta_Z(x) = 1 - \sqrt{x} r_Z(x) $ for $ n = 20 $
and $ b = \kappa^{-2} = 6000 $. Note that the distribution of $ x_i $ as
shown in Fig. \ref{fig:xz} is quite generic for any values of $ n $ and $ b $.
The maxima and minima near both ends ( $ x_1 = 1 $ and $ x_{2n+2} = b $ )
are densely packed even in the logarithm scale, while those at the central
region seem to be evenly distributed in log scale ( i.e., exponentially in
the linear scale ).

\psfigure 5.0in -0.2in {fig:zolo_n6} {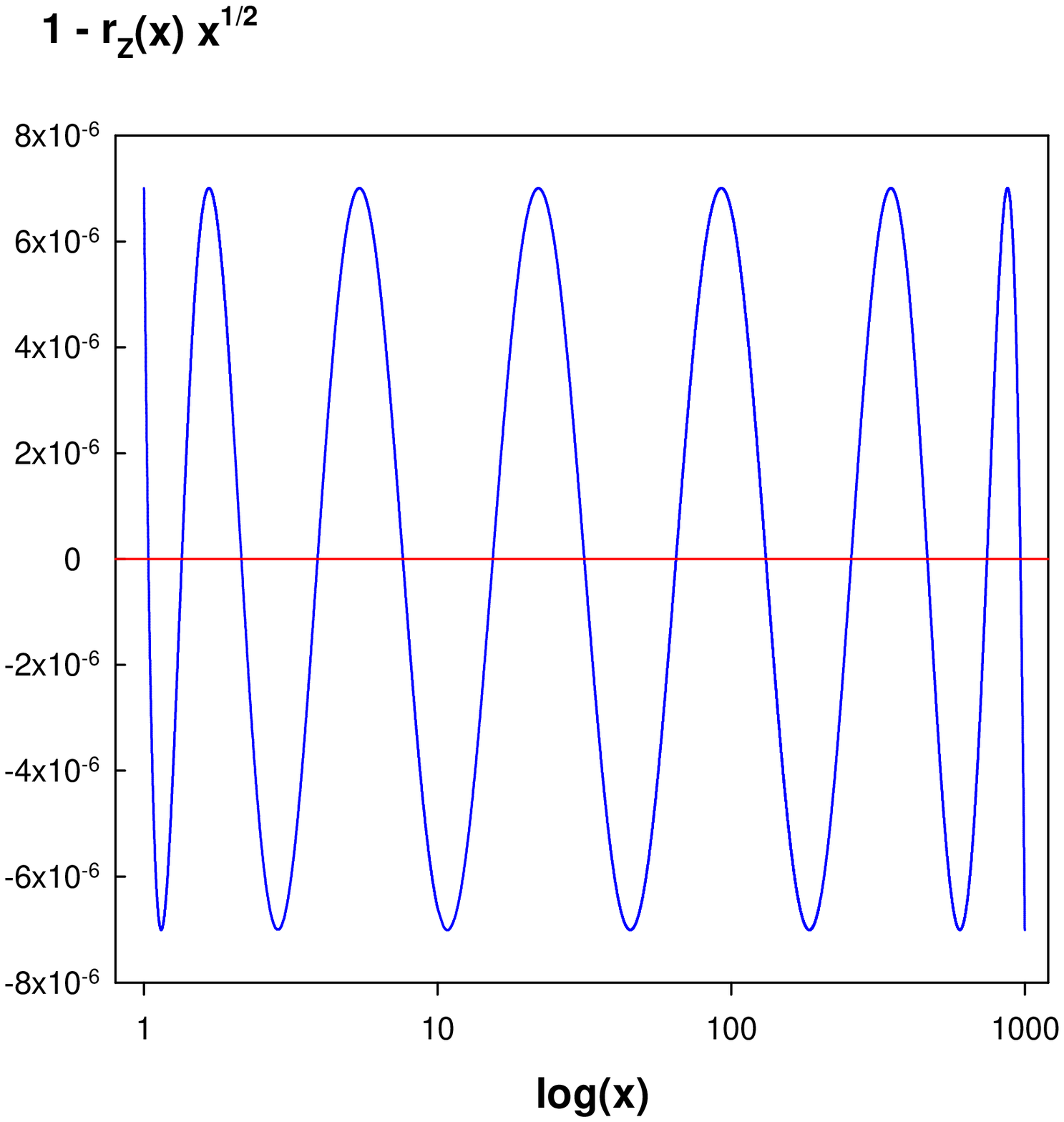} {
The difference $ 1 - \sqrt{x} r_Z(x) $ of Zolotarev optimal rational
approximation (\ref{eq:rZ}) with $ n = 6 $ and $ b = \kappa^{-2} = 1000 $.
Note that there are exactly $ 2n + 2 = 14 $ alternate change
of sign in $ [ 1, 1000 ] $, and the maxima and minima have exactly
the same magnitude. }

\psfigure 5.0in -0.2in {fig:xz} {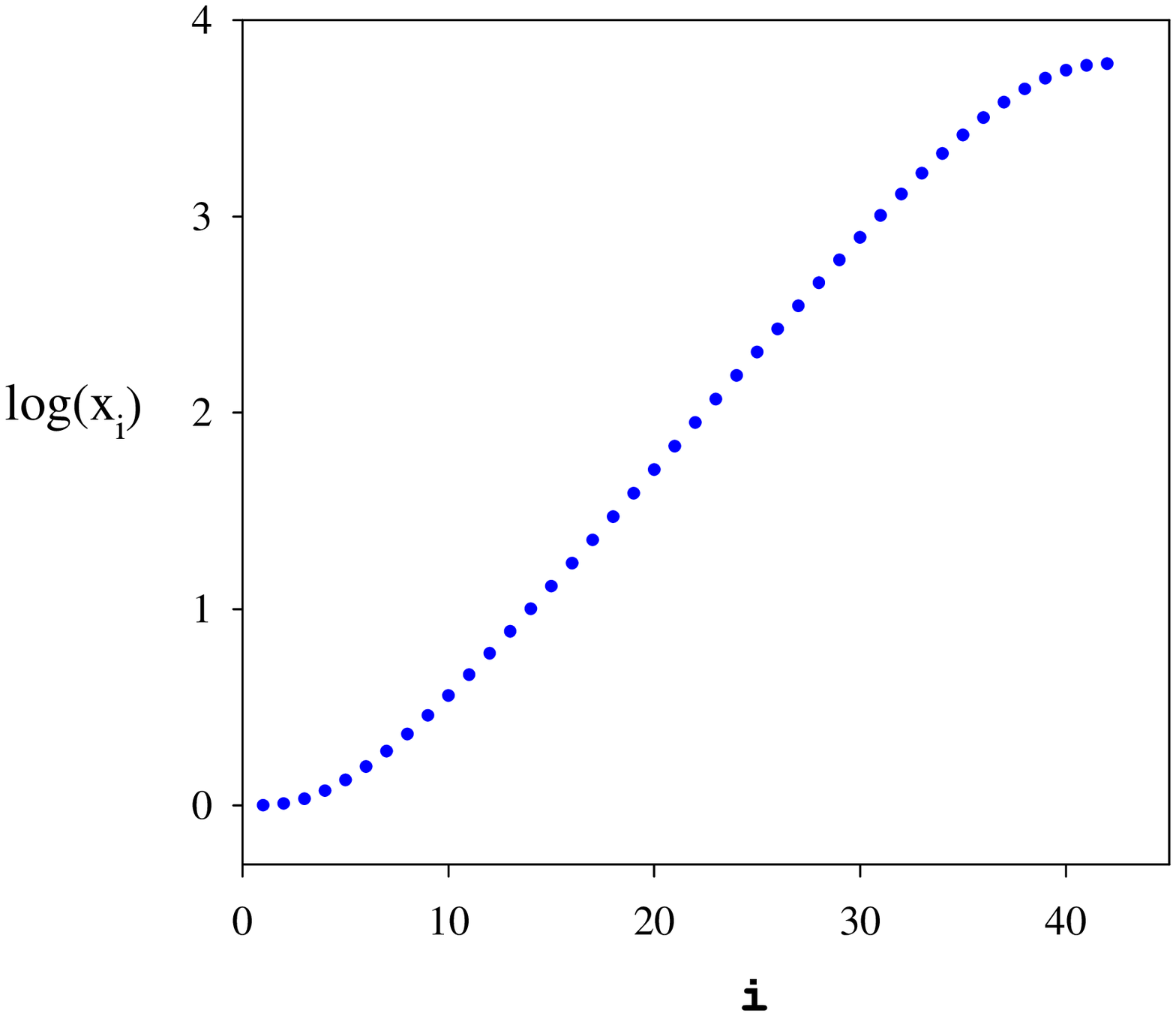} {
The positions $ x_i $ (\ref{eq:xi}) of the maxima and minima of
$ 1 - \sqrt{x} r_Z(x) $ for $ n = 20 $ and $ b = \kappa^{-2} = 6000 $. }

\section{Chebycheff's theorem}

Even though Zolotarev's rational approximation
is optimal among all irreducible rational polynomials of degree $ n $,
with $ \delta(x) = f(x) - g(x) r(x) $ having $ 2 n + 2 $ alternate change
of sign in $ [1,b] $, one may still ask whether there exists an optimal
irreducible rational polynomial of the same degree but its
$ \delta(x) $ has smaller number of alternate change of sign in $ [1,b] $.
Such a possibility is ruled out by Chebycheff's theorem,
which can be shown as follows.

Again, our strategy is to assume the contrary is true, and then show that
it leads to contradiction. Suppose there exists an optimal rational
polynomial $ R(x) = P(x) / Q(x) $ of degree $ n $ with deviation
$ d(R) = d_{min} $, but its $ \Delta(x) \equiv f(x) - g(x) R(x) $
only has $ m = 2n + 1 $ alternate change of sign in $ [1,b] $.
Now assume $ \Delta(x) $ attains its maxima and minima, say,
\bea
\Delta(x) = \Delta_1, -\Delta_2, +\Delta_3, \cdots, +\Delta_{m}
             \hspace{6mm} ( \Delta_i > 0 )
\eea
at consecutive points\footnote{Without loss of generality, we set $ x_1 = 1 $,
and $ x_m = b $.}
\bea
1 = x_1 < x_2 < \cdots < x_{m} = b \ .
\eea
Then the interval $ [1,b] $ can be divided into $ m $ subintervals,
\bea
[1, \xi_1],[ \xi_1, \xi_2], \cdots , [ \xi_{m-1}, b ] \ ,
\hspace{6mm}   x_i  <  \xi_i < x_{i+1}
\eea
such that the following two inequalities can be satisfied alternatively,
\bea
\label{eq:pos}
- d_{min} + \epsilon < & \Delta(x) & \le d_{min} \ , \hspace{4mm}
x \in [ \xi_{2i-2}, \xi_{2i-1} ], \ i=1, \cdots, n+1   \\
\label{eq:neg}
- d_{min} \le & \Delta(x) & < d_{min} - \epsilon \ , \hspace{4mm}
x \in [ \xi_{2i-1}, \xi_{2i} ], \  i = 1, \cdots, n
\eea
where $ \epsilon $ is a positive number which is much less
than $ d_{min} $, and the notations $ \xi_0 = 1 $ and $ \xi_m = b $
have been used.

Next consider the following polynomial of degree $ 2 n = m - 1  $
\bea
\label{eq:S}
S(x) = ( x - \xi_1 ) ( x - \xi_2 ) \cdots ( x - \xi_{m-1})
\eea
which has sign change at $ \xi_i, i=1,\cdots, m-1 $.
Using the fact that $ P(x) $ and $ Q(x) $ have no common factor,
one can find polynomials $ U(x) $ and $ V(x) $ of degree $ n $,
with positive real coefficients, such that
\bea
S(x) = Q(x) U(x) - P(x) V(x) \ .
\eea

Now consider the following irreducible rational polynomial of degree $ n $
\bea
W(x) = \frac{ P(x) + \alpha U(x) }{ Q(x) + \alpha V(x) }
\eea
where $ \alpha $ is a positive real parameter. Then, the difference
\bea
\label{eq:dW}
f(x) - g(x) W(x)
&=& f(x) - g(x) R(x) + g(x) R(x) - g(x) W(x) \nn
&=& \Delta(x) - \alpha g(x) \frac{S(x)}{Q(x)[Q(x) + \alpha V(x)]} \ ,
\eea
is a function of $ \alpha $ which can be chosen to be
some tiny positive numbers such that
the magnitude of the second term on the r.h.s. of (\ref{eq:dW})
is much less than $ d_{min} - \epsilon $ for all
$ x \in [ 1, b ] $, i.e.,
\bea
\label{eq:mu}
0 < \alpha g(x) \frac{|S(x)|}{Q(x)[Q(x) + \alpha V(x)]} \le \mu
\ll  d_{min} - \epsilon \ .
\eea

Since $ \Delta(x) $ and $ S(x) $ both have $ m = 2n + 1 $ alternate
change of sign in $ [ 1, b ] $, and the magnitude of the second term
on r.h.s. of (\ref{eq:dW}) is very small comparing to $ d_{min} - \epsilon $,
it follows that $ f(x) - g(x) W(x) $ also has $ m = 2n + 1 $
alternate change of sign in $ [ 1, b ] $.

Then, using (\ref{eq:pos}), (\ref{eq:S}), (\ref{eq:dW}) and (\ref{eq:mu}),
we obtain
\bea
-d_{min} + \epsilon - \mu < f(x) - g(x) W(x) \le d_{min} - \mu \ ,
\eea
for $ x \in [ \xi_{2i-2}, \xi_{2i-1} ], i=1, \cdots, n+1 $.
Similarly, we have
\bea
-d_{min} + \mu \le f(x) - g(x) W(x) \le d_{min} -\epsilon + \mu \ ,
\eea
for $ x \in [ \xi_{2i-1}, \xi_{2i} ], i=1, \cdots, n $ .
Thus we obtain that $ d(W) = d_{min} - \mu < d_{min} $,
a contradiction to our assumption that $ R(x) $ is the optimal one.
This completes the proof of Chebycheff's theorem.

In general, Chebycheff's theorem asserts that
the optimal irreducible rational polynomial of the form (\ref{eq:rnm})
must satisfy the criterion that the difference
$ \delta(x) = f(x) - g(x) r^{(n,m)}(x) $ has
$ n + m + 2 $ alternate change of sign in the interval $ [ 1, b ] $.

\section{Zolotarev approximation for the overlap}

In this section, we first derive the theoretical error bound for
the the matrix-vector multiplication $ H_w (H_w^2)^{-1/2} Y $
( $ Y $ : any nonzero column vector )
with Zolotarev approximation for $ ( H_w^2 )^{-1/2} $.
Then the numerical values of the error bound are computed,
and listed in Table 1, as well as plotted in Figure 3.
An empirical formula for the error bound is determined from the data.

\subsection{Error bound for $ H_w (H_w^2)^{-1/2}_Z Y $}

To use Zolotarev approximation for the
$ (H_w^2)^{-1/2} $ in the overlap Dirac operator (\ref{eq:Dh}),
one needs to rescale $ H_w $ to $ h_w = H_w / \lambda_{min} $ such that
the eigenvalues of $ h_w^2 $ fall in the interval $ [ 1, b ] $, where
$ b = ( \lambda_{max} / \lambda_{min} )^2 $. Explicitly\footnote{Here we
emphasize that $ d_0 $ (\ref{eq:d0}) can be obtained explicitly, {\it without}
using the condition :
$ \max [ 1 - \sqrt{x} r_{n}(x) ]|_{ 1 \le x \le b } =
- \min [ 1 - \sqrt{x} r_{n}(x) ]|_{ 1 \le x \le b } $.}
\bea
\label{eq:zolo}
  \frac{1}{\sqrt{H_w^2}}
\simeq \frac{d_0}{\lambda_{min}}
  \prod_{l=1}^{n} \frac{ h_w^2 + c_{2l} }{ h_w^2 + c_{2l-1} }
=  \frac{1}{\lambda_{min}} ( h_w^2 + c_{2n} )
   \sum_{l=1}^{n} \frac{ b_l }{ h_w^2 + c_{2l-1} } \equiv
   (H_w^2)^{-1/2}_Z
\eea
where
\bea
\label{eq:d0}
d_0 &=& \frac{2 \lambda }{1+ \lambda}
        \prod_{l=1}^n \frac{1+c_{2l-1}}{1+c_{2l}} \\
b_l &=& d_0 \frac{ \prod_{i=1}^{n-1} ( c_{2i} - c_{2l-1} ) }
             { \prod_{i=1, i \ne l}^{n} ( c_{2i-1} - c_{2l-1} ) } \ .
\eea

First, we consider the multiplication of $ H_w (H_w^2)^{-1/2} $ to a
nonzero column vector $ Y $
\bea
\label{eq:mult_Y}
H_w \left( \frac{1}{\sqrt{H_w^2}} \right) Y \simeq
  h_w ( h_w^2 + c_{2n} )
  \sum_{l=1}^{n} \frac{ b_l }{ h_w^2 + c_{2l-1} } Y
= h_w ( h_w^2 + c_{2n} )  \sum_{l=1}^{n} b_l Z_l
\eea
where the last summation can be evaluated by invoking a conjugate
gradient process to the linear systems
\bea
\label{eq:inner_CG}
( h_w^2 + c_{2l-1} ) Z_l = Y, \hspace{4mm} l = 1, \cdots, n \ .
\eea
In order to improve the accuracy of the rational approximation as well as
to reduce the number of iterations in the conjugate gradient loop,
it is advantageous to narrow the interval $ [ 1, b ] $ by projecting out
the largest and some low-lying eigenmodes of $ H_w^2 $.
Denoting these eigenmodes by
\bea
\label{eq:eigen}
H_w u_j = \lambda_j u_j, \hspace{4mm} j = 1, \cdots, k ,
\eea
then we project the linear systems (\ref{eq:inner_CG}) to the
complement of the vector space spanned by these eigenmodes
\bea
\label{eq:inner_CG1}
( h_w^2 + c_{2l-1} ) \bar{Z_l} = \bar{Y}
\equiv ( 1 - \sum_{j=1}^k u_j u_j^{\dagger} ) Y \ ,
\hspace{4mm} l = 1, \cdots, n \ .
\eea

In the set of projected eigenvalues of $ H_w^2 $,
$ \{ \lambda_j^2 , j = 1, \cdots, k \} $,
we use $ \lambda_{max}^2 $ and $ \lambda_{min}^2 $ to denote
the least upper bound and the greatest lower bound
for the eigenvalues of $ \bar{H}_w^2 $, where
\BAN
\bar{H}_w = H_w - \sum_{j=1}^k \lambda_j u_j u_j^{\dagger} \ .
\EAN
Then the eigenvalues of
\BAN
h_w^2 = \bar{H}_w^2 / \lambda_{min}^2
\EAN
fall into the interval $ (1,b) $, $ b = ( \lambda_{max}/\lambda_{min} )^2 $.

Now the matrix-vector multiplication (\ref{eq:mult_Y})
can be expressed in terms of the projected eigenmodes (\ref{eq:eigen})
plus the solution obtained from the conjugate gradient loop
(\ref{eq:inner_CG1}) in the complementary vector space, i.e.,
\bea
\label{eq:SS}
H_w \frac{1}{\sqrt{H_w^2}} Y \simeq
\frac{1}{\lambda_{min}} \bar{H}_w ( h_w^2 + c_{2n} )
                        \sum_{l=1}^{n} b_l \bar{Z}_l
  + \sum_{j=1}^k  \frac{\lambda_j}{\sqrt{\lambda_j^2}}
                   u_j u_j^{\dagger} Y \equiv S
\eea
Then the error of $ S $ can be measured in terms of
\bea
\label{eq:sigma}
\sigma = \frac{ | S^{\dagger} S - Y^{\dagger} Y | }{  Y^{\dagger} Y } \ ,
\eea
which is zero if (\ref{eq:SS}) is exact.
Now assuming the errors due to the conjugate gradient
and the projected eigenmodes are negligible compared with that due
to the Zolotarev approximation of $ (h_w^2)^{-1/2} $,
then we can derive the theoretical upper bound of $ \sigma $ as follows.

First, we rewrite $ S $ (\ref{eq:SS}) as
\bea
\label{eq:SSS}
S = \sum_{j=1}^k \frac{\lambda_j}{\sqrt{\lambda_j^2}} u_j u_j^{\dagger} Y +
\sum_{j=k+1}^N
\mbox{sign}(\lambda_j) \sqrt{x_j} r_Z(x_j) u_j u_j^{\dagger} Y \ ,
\hspace{4mm} x_j \equiv \left( \frac{\lambda_j}{\lambda_{min}} \right)^2
\eea
where $ \lambda_j, u_j, j=k+1, \cdots, N $ denote the eigenvalues and
eigenfunctions of $ \bar{H}_w $.
Then it is straightforward to derive
\bea
  S^{\dagger} S - Y^{\dagger} Y
= Y^{\dagger} \left(
  \sum_{j=k+1}^N [ ( \sqrt{x_j} r_Z(x_j) )^2 - 1 ] u_j u_j^{\dagger}
  \right) Y \ ,
\eea
where the orthonormality ( $ u_i^{\dagger} u_j = \delta_{ij} $ )
has been used. Thus it follows that
\bea
| S^{\dagger} S - Y^{\dagger} Y |
& \le & \max_{ k+1 \le j \le N } | ( \sqrt{x_j} r_Z(x_j) )^2 - 1 | \
Y^{\dagger} \left( \sum_{l=k+1}^N u_l u_l^{\dagger} \right) Y \nn
& < & \max_{ k+1 \le j \le N } | ( \sqrt{x_j} r_Z(x_j) )^2 - 1 | \
      Y^{\dagger} Y
\eea
where the inequality ( due to the completeness relation
$  \sum_{l=1}^N u_l u_l^{\dagger} = \Id $ )
\BAN
    Y^{\dagger} \left( \sum_{l=k+1}^N u_l u_l^{\dagger} \right) Y
<   Y^{\dagger} Y \ ,  \hspace{4mm} ( k > 0 )
\EAN
has been used.
Then the theoretical upper bound of (\ref{eq:sigma}) is
\bea
\label{eq:sigma_bound}
\sigma < \max_{ k+1 \le j \le N } | ( \sqrt{x_j} r_Z(x_j) )^2 - 1 |
\simeq 2 \max_{ k+1 \le j \le N } | \sqrt{x_j} r_Z(x_j) - 1 | \ .
\eea
Thus $ \sigma $ is less than two times of the maximum of
$ | 1 - \sqrt{x_j} r_Z(x_j) | $ for all eigenvalues of $ h_w^2 $.
It follows that $ \sigma $ must be less than two times of
the maximum of $ |1-\sqrt{x} r_Z(x)| $ for all $ x \in [1,b] $, i.e.,
\bea
\label{eq:sigma_th_bd}
\sigma = \frac{ | S^{\dagger} S - Y^{\dagger} Y | }{  Y^{\dagger} Y }
       < \frac{2(1-\lambda)}{1+\lambda} \equiv 2 d_Z(n,b) \ ,
\eea
where $ \lambda $ is defined in (\ref{eq:lambda}).
The inequality (\ref{eq:sigma_th_bd}) is one of the main results of
this paper.

A remarkable feature of (\ref{eq:sigma_th_bd}) is that it holds
for any nonzero column vector $ Y $. Then one immediately sees that
the deviation (\ref{eq:Delta_Z}) which measures the
chiral symmetry breaking due to Zolotarev approximation also
has the same theoretical upper bound,
\bea
\label{eq:DZB}
\Delta_Z < 2 d_Z(n,b) \ .
\eea
Thus $ 2 d_Z(n,b) $ not only serves as the theoretical error bound
for the matrix-vector multiplication $ H_w ( H_w^2 )^{-1/2}_Z Y $, but
also provides the upper bound for the chiral symmetry breaking due to
Zolotarev approximation. Note that $ d_Z(n,b) $ does not depend on the
lattice size explicitly.
Therefore, by choosing the proper values of $ n $ and $ b $
( i.e., by projecting the high and low-lying eigenmodes of $ H_w^2 $ ),
one can practically preserve the exact chiral symmetry of
the overlap Dirac operator to very high precision, for
any gauge configurations on a finite lattice.

Moreover, for any gauge configuration, it is unlikely
that any of the eigenvalues of $ h_w^2 $ would coincide with one of
those $ 2 n + 2 $ positions with maximum deviation (\ref{eq:xi}),
thus one usually obtains a $ \sigma $ much
smaller than the theoretical error bound $ 2 d_Z(n,b) $.

In Table \ref{table:sigma}, we list the values of $ \sigma $
(\ref{eq:sigma}) for several gauge configurations on
three different lattices respectively, along with the
theoretical error bound $ 2 d_Z(n,b) $.
It is clear that $ \sigma $ is always much
smaller than the theoretical error bound.
Note that for each configuration,
we only show the largest $ \sigma $ among a set of several hundred
$ \sigma $ values, each is computed with a different $ Y $ at every
iteration of the outer CG loop.
The average value of $ \sigma $ is usually less than $ 1/5 $ of the
largest one listed in Table \ref{table:sigma}.
Details of our computation are described in \cite{Chiu:2002xm}.
Here we have set the stopping criterion for the inner and outer
conjugate gradient loops at $ \epsilon = 10^{-11} $,
and the error of the projected eigenmodes around $ 10^{-13} $.
It is remarkable that $ \sigma $ turns out to be much less than
$ \epsilon $, in contrast to one's naive expectation.
In other words, even if the precision of
the overlap Dirac quark propagator is only up to $ 10^{-11} $,
its exact chiral symmetry can attain $ \Delta_Z < 10^{-12} $.

{\footnotesize
\begin{table}
\begin{center}
\begin{tabular}{|c|c|c|c|c|c|c|c|}
\hline
lattice size & $ \beta $  & $ \lambda_{min} $  & $ \lambda_{max} $ & $ b $ &
$ n $ & $ \sigma $ & $ 2 d_Z(n,b) $       \\
\hline
\hline
$ 16^3 \times 32 $ & $ 6.0 $ & $ 0.1731 $ & $ 6.258 $ & 1307.00 & $ 12 $ &
$ 6.0 \times 10^{-12} $ & $ 1.4 \times 10^{-10} $ \\
\hline
$ 16^3 \times 32 $ & $ 6.0 $ & $ 0.1943 $ & $ 6.260 $ & 1038.01 & $ 12 $ &
$ 7.0 \times 10^{-12} $ & $ 7.5 \times 10^{-11} $ \\
\hline
$ 16^3 \times 32 $ & $ 6.0 $ & $ 0.1767 $ & $ 6.261 $ & 1255.49 & $ 12 $ &
$ 8.0 \times 10^{-12} $ & $ 1.2 \times 10^{-10} $ \\
\hline
$ 16^3 \times 32 $ & $ 6.0 $ & $ 0.1955 $ & $ 6.260 $ & 1025.31 & $ 12 $ &
$ 5.0 \times 10^{-12} $ & $ 7.2 \times 10^{-11} $ \\
\hline
$ 12^3 \times 24 $ & $ 5.8 $ & $ 0.1176 $ & $ 6.210 $ & 2788.49 & $ 16 $ &
$ 7.0 \times 10^{-14} $ & $ 4.8 \times 10^{-13} $ \\
\hline
$ 12^3 \times 24 $ & $ 5.8 $ & $ 0.1285 $ & $ 6.211 $ & 2336.24 & $ 16 $ &
$ 6.6 \times 10^{-14} $ & $ 3.2 \times 10^{-13} $ \\
\hline
$ 12^3 \times 24 $ & $ 5.8 $ & $ 0.0988 $ & $ 6.206 $ & 3945.57 & $ 16 $ &
$ 2.0 \times 10^{-13} $ & $ 1.3 \times 10^{-12} $ \\
\hline
$ 12^3 \times 24 $ & $ 5.8 $ & $ 0.1415 $ & $ 6.213 $ & 1927.92 & $ 16 $ &
$ 5.5 \times 10^{-14} $ & $ 1.6 \times 10^{-13} $ \\
\hline
$ 10^3 \times 24 $ & $ 5.8 $ & $ 0.1242 $ & $ 6.214 $ & 2503.22 & $ 16 $ &
$ 1.6 \times 10^{-13} $ & $ 3.4 \times 10^{-13} $ \\
\hline
$ 10^3 \times 24 $ & $ 5.8 $ & $ 0.1381 $ & $ 6.209 $ & 2021.42 & $ 16 $ &
$ 6.1 \times 10^{-14} $ & $ 2.2 \times 10^{-13} $ \\
\hline
$ 10^3 \times 24 $ & $ 5.8 $ & $ 0.1376 $ & $ 6.204 $ & 2032.86 & $ 16 $ &
$ 5.5 \times 10^{-14} $ & $ 2.6 \times 10^{-13} $ \\
\hline
$ 10^3 \times 24 $ & $ 5.8 $ & $ 0.1181 $ & $ 6.204 $ & 2759.59 & $ 16 $ &
$ 4.4 \times 10^{-14} $ & $ 5.0 \times 10^{-13} $ \\
\hline
\end{tabular}
\end{center}
\caption{
The error $ \sigma $ (\ref{eq:sigma}) of the matrix-vector
multiplication $ H_w (H_w^2)^{-1/2}_Z Y $,
for several gauge configurations on three different lattices.
Evidently, each $ \sigma $ is quite smaller than the corresponding
theoretical error bound $ 2 d_Z(n,b) = 2 ( 1-\lambda )/(1+\lambda) $.
}
\label{table:sigma}
\end{table}
}

\subsection{An empirical formula for the error bound}

In Table \ref{table:drZ}, we list the values of
$ d_Z(n,b)=(1-\lambda)/(1+\lambda) $
of Zolotarev optimal rational appproximation (\ref{eq:rZ}),
for $ n \simeq 10 - 20 $ and $ b \simeq 10 - 10^6 $.
We can use Table \ref{table:drZ} to determine
how many Zolotarev terms ( $ n $ ) is needed in order to
attain one's desired accuracy, after the highest and the
low-lying eigenmodes are projected out and
$ b = ( \lambda_{max} / \lambda_{min} )^2 $ has been obtained.
Conversely, for a fixed number of Zolotarev terms, say $ n $,
one can use Table \ref{table:drZ} to determine what ranges of high and
low-lying eigenmodes of $ H_w^2 $ should be projected in order to
attain one's desired accuracy.

{\footnotesize
\begin{table}
\begin{center}
\begin{tabular}{|c|c|c|c|c|c|c|}
\hline
$ b $
& \multicolumn{6}{c|}{ $ n $ }  \\
\hline
       &   10  &  12   &   14  &  16  &  18  &  20  \\
\hline
\hline
$ 10 $ & $ 4.8 \times 10^{-18} $ & $ 1.9 \times 10^{-21} $
       & $ 7.2 \times 10^{-25} $ & $ 2.8 \times 10^{-28} $
       & $ 1.1 \times 10^{-31} $ & $ 4.1 \times 10^{-35} $  \\
\hline
$ 50 $ & $ 1.3 \times 10^{-13} $ & $ 3.5 \times 10^{-16} $
       & $ 9.5 \times 10^{-19} $ & $ 2.6 \times 10^{-21} $
       & $ 6.9 \times 10^{-24} $ & $ 1.9 \times 10^{-26} $  \\
\hline
$ 100 $ & $ 2.5 \times 10^{-12} $ & $ 1.2 \times 10^{-14} $
        & $ 5.5 \times 10^{-17} $ & $ 2.6 \times 10^{-19} $
        & $ 1.2 \times 10^{-21} $ & $ 5.8 \times 10^{-24} $  \\
\hline
$ 500 $ & $ 3.8 \times 10^{-10} $ & $ 4.8 \times 10^{-12} $
        & $ 5.9 \times 10^{-14} $ & $ 7.3 \times 10^{-16} $
        & $ 9.0 \times 10^{-18} $ & $ 1.1 \times 10^{-19} $  \\
\hline
$ 1000 $ & $ 2.0 \times 10^{-9}  $ & $ 3.4 \times 10^{-11} $
                  & $ 5.8 \times 10^{-13} $ & $ 9.8 \times 10^{-15} $
                  & $ 1.7 \times 10^{-16} $ & $ 2.8 \times 10^{-18} $  \\
\hline
$ 2000 $ & $ 8.4 \times 10^{-9}  $ & $ 1.9 \times 10^{-10} $
                  & $ 4.2 \times 10^{-12} $ & $ 9.3 \times 10^{-14} $
                  & $ 2.1 \times 10^{-15} $ & $ 4.6 \times 10^{-17} $  \\
\hline
$ 3000 $ & $ 1.8 \times 10^{-8}  $ & $ 4.6 \times 10^{-10} $
                  & $ 1.2 \times 10^{-11} $ & $ 3.0 \times 10^{-13} $
                  & $ 7.7 \times 10^{-15} $ & $ 2.0 \times 10^{-16} $  \\
\hline
$ 4000 $ & $ 2.9 \times 10^{-8}  $ & $ 8.3 \times 10^{-10} $
                  & $ 2.3 \times 10^{-11} $ & $ 6.6 \times 10^{-13} $
                  & $ 1.9 \times 10^{-14} $ & $ 5.3 \times 10^{-16} $  \\
\hline
$ 5000 $ & $ 4.3 \times 10^{-8}  $ & $ 1.3 \times 10^{-9}  $
                  & $ 3.9 \times 10^{-11} $ & $ 1.2 \times 10^{-12} $
                  & $ 3.6 \times 10^{-14} $ & $ 1.1 \times 10^{-15} $  \\
\hline
$ 6000 $ & $ 5.7 \times 10^{-8}  $ & $ 1.8 \times 10^{-9}  $
                  & $ 5.8 \times 10^{-11} $ & $ 1.9 \times 10^{-12} $
                  & $ 6.0 \times 10^{-14} $ & $ 1.9 \times 10^{-15} $  \\
\hline
$ 7000 $ & $ 7.2 \times 10^{-8}  $ & $ 2.4 \times 10^{-9}  $
                  & $ 8.1 \times 10^{-11} $ & $ 2.7 \times 10^{-12} $
                  & $ 9.1 \times 10^{-14} $ & $ 3.1 \times 10^{-15} $  \\
\hline
$ 8000 $ & $ 8.9 \times 10^{-8}  $ & $ 3.1 \times 10^{-9}  $
                  & $ 1.1 \times 10^{-10} $ & $ 3.7 \times 10^{-12} $
                  & $ 1.3 \times 10^{-13} $ & $ 4.5 \times 10^{-15} $  \\
\hline
$ 9000 $ & $ 1.1 \times 10^{-7}  $ & $ 3.8 \times 10^{-9}  $
                  & $ 1.4 \times 10^{-10} $ & $ 4.9 \times 10^{-12} $
                  & $ 1.8 \times 10^{-13} $ & $ 6.4 \times 10^{-15} $  \\
\hline
$ 1 \times 10^4 $ & $ 1.2 \times 10^{-7}  $ & $ 4.6 \times 10^{-9}  $
                  & $ 1.7 \times 10^{-10} $ & $ 6.3 \times 10^{-12} $
                  & $ 2.3 \times 10^{-13} $ & $ 8.6 \times 10^{-15} $  \\
\hline
$ 5 \times 10^4 $ & $ 9.5 \times 10^{-7}  $ & $ 5.2 \times 10^{-8}  $
                  & $ 2.9 \times 10^{-9}  $ & $ 1.6 \times 10^{-10} $
                  & $ 8.6 \times 10^{-12} $ & $ 4.7 \times 10^{-13} $  \\
\hline
$ 1 \times 10^5 $ & $ 2.0 \times 10^{-6}  $ & $ 1.3 \times 10^{-7}  $
                  & $ 8.0 \times 10^{-9}  $ & $ 5.0 \times 10^{-10} $
                  & $ 3.2 \times 10^{-11} $ & $ 2.0 \times 10^{-12} $  \\
\hline
$ 5 \times 10^5 $ & $ 8.7 \times 10^{-6}  $ & $ 7.3 \times 10^{-7}  $
                  & $ 6.1 \times 10^{-8}  $ & $ 5.0 \times 10^{-9} $
                  & $ 4.2 \times 10^{-10} $ & $ 3.5 \times 10^{-11} $  \\
\hline
$ 1 \times 10^6 $ & $ 1.5 \times 10^{-5}  $ & $ 1.4 \times 10^{-6}  $
                  & $ 1.3 \times 10^{-7}  $ & $ 1.2 \times 10^{-8} $
                  & $ 1.1 \times 10^{-9}  $ & $ 1.0 \times 10^{-10} $  \\
\hline
\end{tabular}
\end{center}
\caption{
The deviation
$ d_Z(n,b) = \max | 1 - \sqrt{x} r^{(n,n)}_Z(x) |
= ( 1-\lambda )/(1+\lambda) $
of the Zolotarev optimal rational approximation (\ref{eq:rZ}),
versus the upper bound $ b $ of $ x \in [ 1, b ] $,
and the degree $ n $ of the rational polynomial.}
\label{table:drZ}
\end{table}
}

In Fig. \ref{fig:dev_zolo}, we plot
$ d_Z(n,b) = \max | 1 - \sqrt{x} r^{(n,n)}_Z(x) | = (1-\lambda)/(1+\lambda) $
versus the degree $ n $,
for different values of $ b $ ranging from $ 10^3 - 10^6 $.
For any fixed value of $ b $, the error bound converges exponentially
with respect to $ n $, and it is well fitted by
\bea
\label{eq:eZ}
e_Z(n,b) =  A(b) \exp\{ - c(b) n \} \ ,
\eea
as indicated by the solid lines shown in Fig. \ref{fig:dev_zolo}.
The parameters $ c(b) $ and $ A(b) $ are determined in Figs. \ref{fig:cb}
and \ref{fig:Ab} respectively,
\bea
\label{eq:cb}
c(b) &=& 9.17(10) \ \mbox{ln} (b)^{-0.774(5)} \\
\label{eq:Ab}
A(b) &=& 0.465(9) \ \mbox{ln} (b)^{0.596(9)}
\eea

With the empiraical formula (\ref{eq:eZ}), one can estimate the theoretical
upper bound of $ \sigma $ or $ \Delta_Z $ more conveniently,
without evaluating ellitpic functions at all.

\psfigure 5.0in -0.2in {fig:dev_zolo} {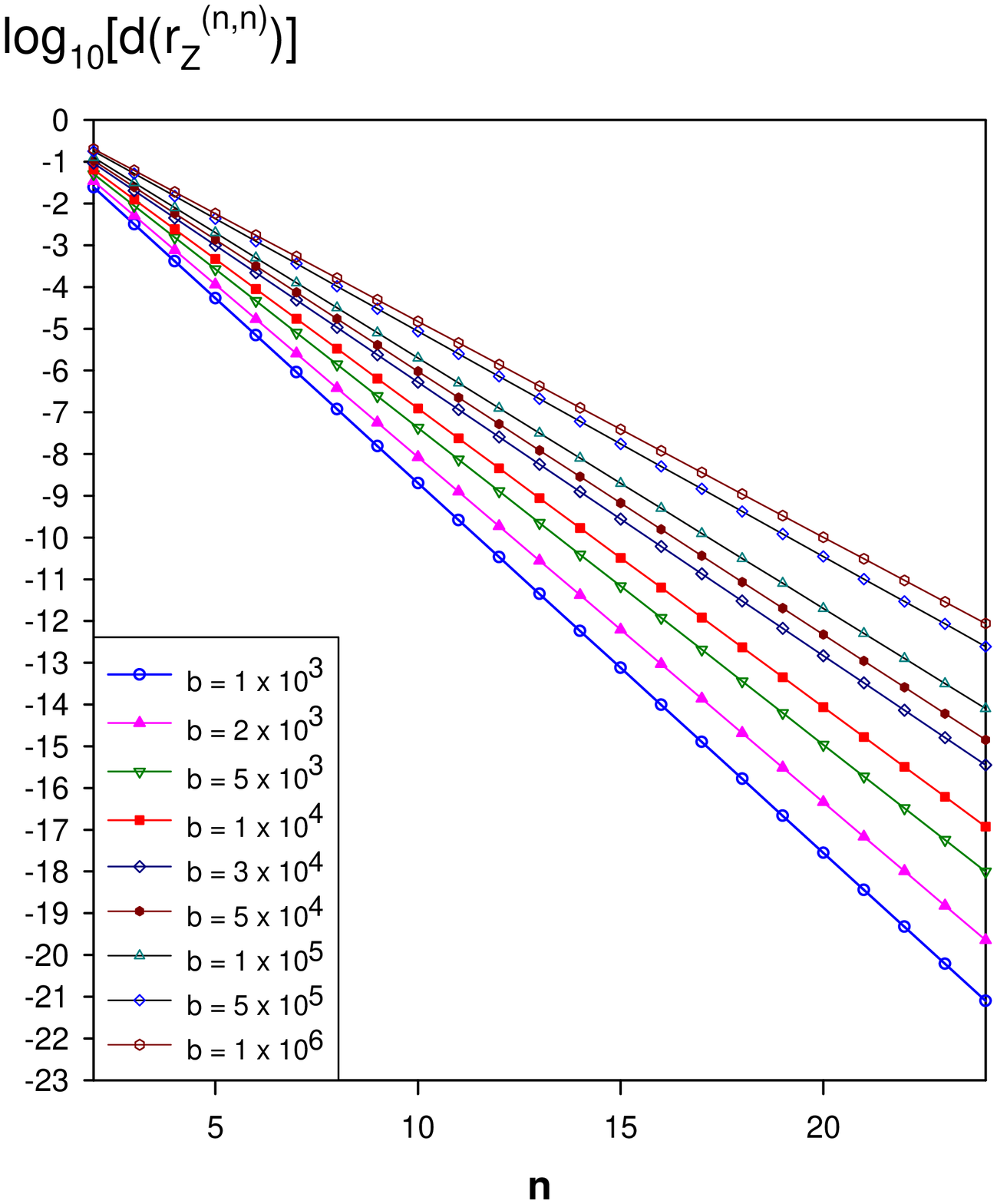} {
The error bound of Zolotarev optimal rational approximation
for the inverse square root function $ x^{-1/2}, x \in [ 1, b ] $,
versus the degree $ n $ of the rational polynomial,
and for several different values of $ b $ ranging from $ 10^3 $
to $ 10^5 $. For any fixed value of $ b $, the error bound converges
exponentially with respect to $ n $, and is well fitted by
$ A(b) \exp\{ - c(b) n \} $ with
$ c(b) =  9.17(10) \mbox{ln}(b)^{-0.774(5)} $
and $ A(b) = 0.465(9) \mbox{ln}(b)^{0.596(9)} $.}

\psfigure 5.0in -0.2in {fig:cb} {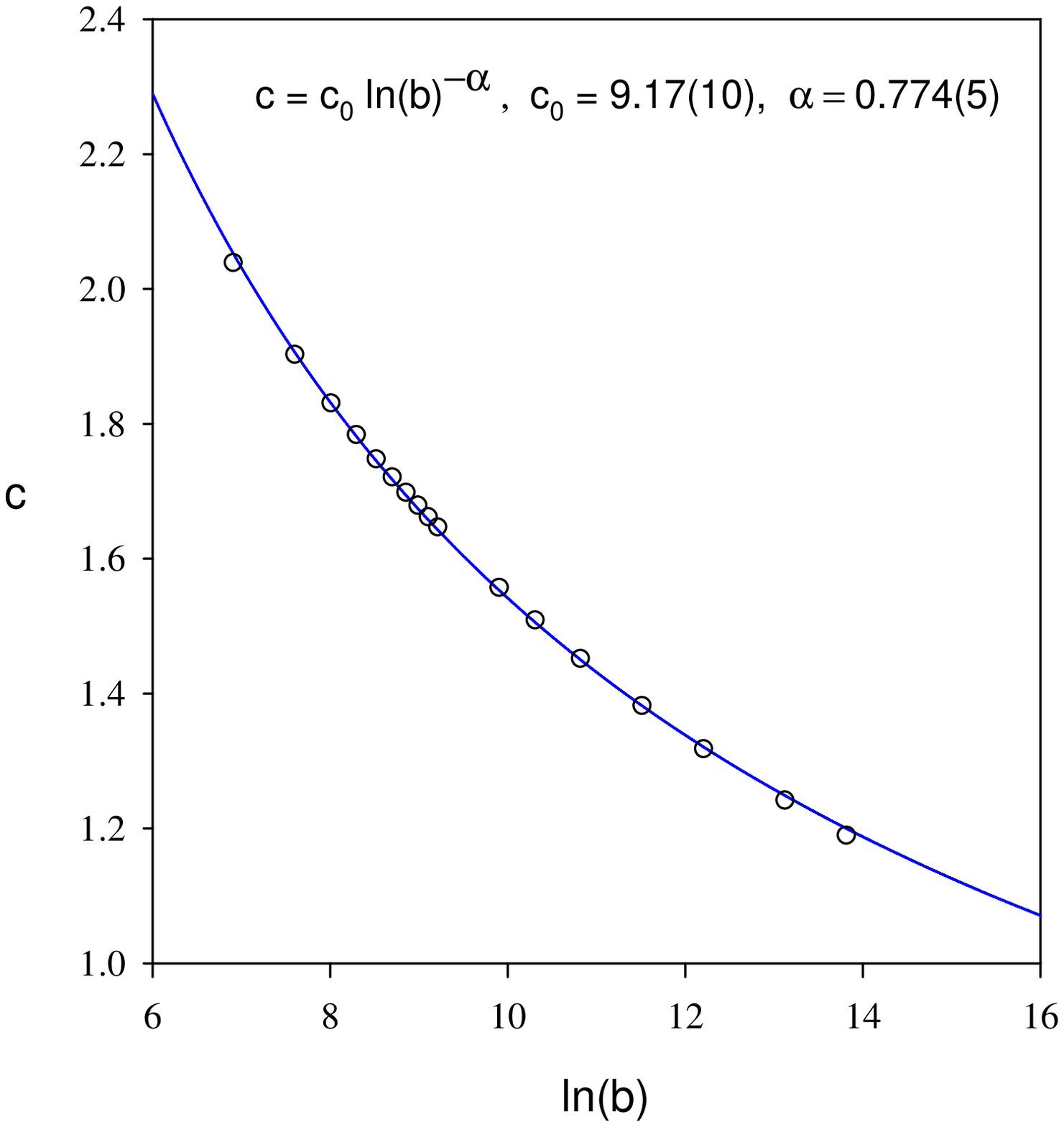} {
Determination of the parameters of $ c(b) $ }

\psfigure 5.0in -0.2in {fig:Ab} {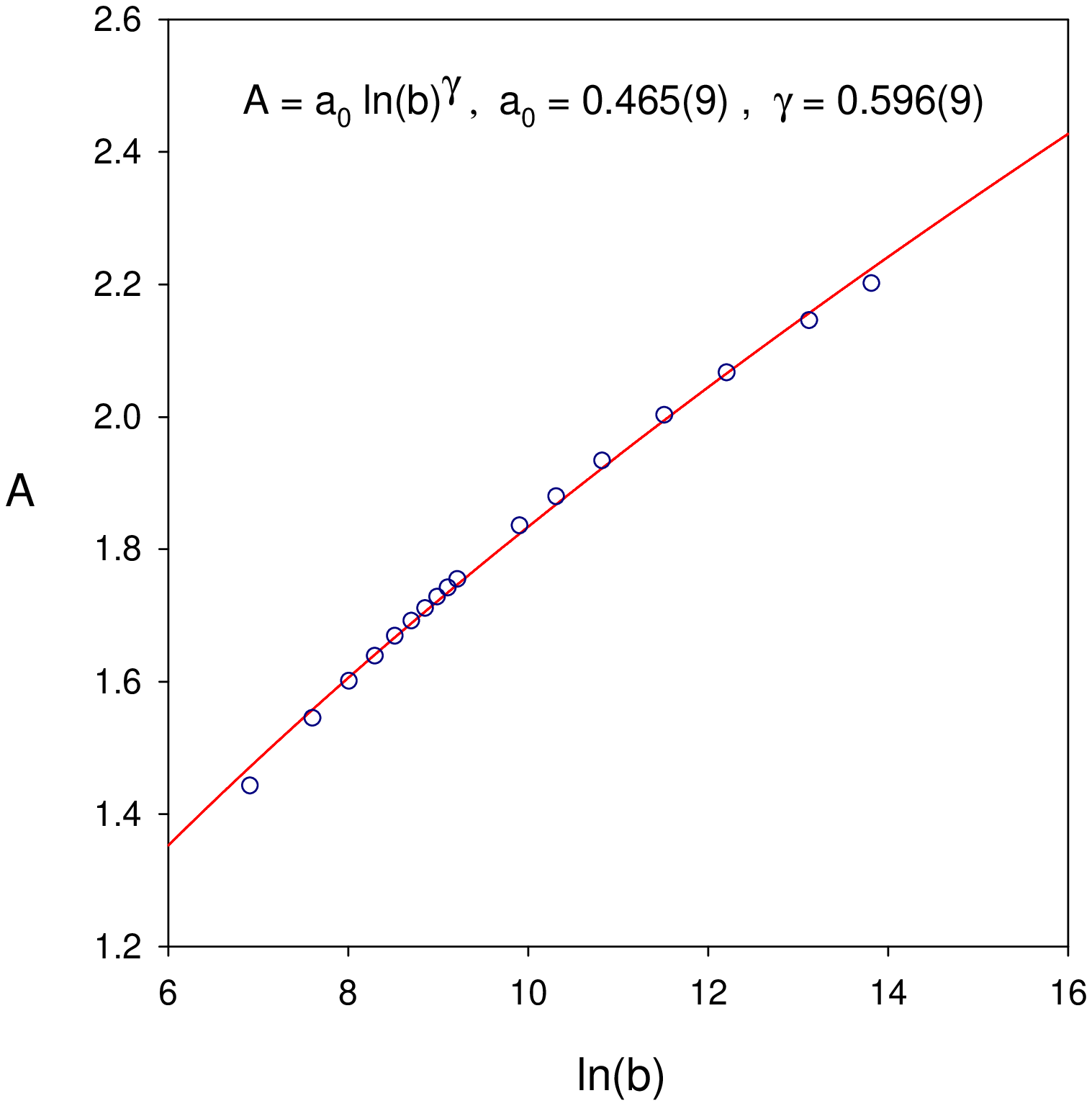} {
Determination of the parameters of $ A(b) $ }

\subsection{Zolotarev rational polynomial of the form $ r^{(n-1,n)} $}

At this point, if one recalls Chebycheff's therorem for $ r^{(n,m)} $,
one may ask whether there exists optimal rational approximation
to $ x^{-1/2} $, which has the form $ r^{(n-1,n)} $,
with $ 1 - \sqrt{x} r^{(n-1,n)}_Z(x) $
having $ 2 n + 1 $ alternate change of sign in $ [1,b] $.
Note that in the second principal $n$-th degree
transformation (\ref{eq:transf}),
$ \mbox{sn}( x/M ; \lambda ) $ has periods $ 2 K/M $ and $ 2 i K'/[(2n+1)M] $.
Thus there must exist a similar transformation in which
$ \mbox{sn}( x/M ; \lambda ) $ has periods $ 2 K/M $ and $ 2 i K'/2nM $.
Explicitly, it reads
\bea
\sqrt{ x \left( \frac{u}{m};\Lambda \right) } =
\sqrt{ x (u;\kappa) } \frac{1}{m}
  \frac{ \prod_{l=1}^{n-1} [ 1 + x(u;\kappa)/{C}_{2l} ] }
       { \prod_{l=1}^{n} [ 1+ x(u;\kappa)/{C}_{2l-1} ] }
\eea
where
\bea
\label{eq:Lambda}
\Lambda &=& \prod_{l=1}^{2n}
\frac{\Theta^2 \left(\frac{2lK'}{2n};\kappa' \right)}
     {\Theta^2 \left(\frac{(2l-1)K'}{2n};\kappa' \right)} \ , \\
\label{eq:m}
m &=& \frac
{\prod_{l=1}^{n} \mbox{sn}^2 \left(\frac{(2l-1)K'}{2n};\kappa' \right) }
{\prod_{l=1}^{n-1} \mbox{sn}^2 \left(\frac{2lK'}{2n};\kappa' \right) } \ , \\
\label{eq:cl'}
{C}_l &=& \frac{\mbox{sn}^2(\frac{lK'}{2n}; \kappa' ) }
             {1-\mbox{sn}^2(\frac{lK'}{2n}; \kappa' )},
\eea

Then the Zolotarev optimal rational polynomial of the form $ r^{(n-1,n)} $ is
\bea
\label{eq:rZ'}
r^{(n-1,n)}_Z (x) = \frac{2 \Lambda}{ 1 + \Lambda } \frac{1}{m}
\frac{ \prod_{l=1}^{n-1} ( 1 + x/{C}_{2l} ) }
     { \prod_{l=1}^{n} ( 1 + x/{C}_{2l-1} ) }
\eea

The difference
$ 1 - \sqrt{x} r^{(n-1,n)}_Z(x) $
has $ 2n + 1 $ alternate change of sign in $ [1,b] $,
( $ b = \kappa^{-2} $ ),
and it attains its maxima and minima alternatively as
\BAN
  \frac{1-\Lambda}{1+\Lambda}, -\frac{1-\Lambda}{1+\Lambda}, \cdots,
 -\frac{1-\Lambda}{1+\Lambda},  \frac{1-\Lambda}{1+\Lambda}
\EAN
at $ 2n + 1 $ successive points $ x \in [ 1, b ] $ :
\BAN
1,  \frac{1}
        {1-\kappa'^2 \mbox{sn}^2 \left(\frac{K'}{2n}; \kappa' \right)},
\cdots,
\frac{1}
{ 1- \kappa'^2 \mbox{sn}^2 \left( \frac{(2n-1)K'}{2n}; \kappa' \right)},
\frac{1}{\kappa^2} \ .
\EAN

Obviously, for any given $ n $ and $ b $, the deviation of
$ r^{(n-1,n)}_Z $ is larger than that of $ r^{(n,n)}_Z $, i.e.,
\bea
d( r^{(n-1,n)}_Z ) = \frac{1- \Lambda}{1+\Lambda} >
           \frac{1- \lambda}{1+\lambda} = d( r^{(n,n)}_Z ) \ ,
\eea
and for most cases,
\bea
d( r^{(n-1,n)}_Z ) \simeq 2.5 \times d( r^{(n,n)}_Z )
\eea

In Table \ref{table:drZ'}, we list the deviation
$ d(r^{(n-1,n)}_Z)=(1 - \Lambda)/(1 + \Lambda) $
of the Zolotarev optimal rational appproximation (\ref{eq:rZ'}),
versus the degree $ n $,
and the upper bound $ b $ of $ x \in [ 1, b ] $.
Comparing the corresponding entries ( with the same $ b $ and $ n $ )
in Table \ref{table:drZ'} and Table \ref{table:drZ},
one immediately sees that $ d(r^{(n-1,n)}_Z) \simeq 2.5 \times d(r^{(n,n)}_Z) $.

If one uses (\ref{eq:rZ'}) to approximate the inverse square root
of $ H_w^2 $ in the overlap Dirac operator, then one has
\bea
\label{eq:zol}
\frac{1}{\sqrt{H_w^2}}
\simeq \frac{ D_0}{\lambda_{min}}
       \frac{ \prod_{l=1}^{n-1} ( h_w^2 + C_{2l} ) }
            { \prod_{l=1}^{n} ( h_w^2 + C_{2l-1} ) }
= \frac{D_0}{\lambda_{min}} \sum_{l=1}^{n} \frac{ B_l }{ h_w^2 + C_{2l-1} }
\eea
where
\bea
\label{eq:D0}
D_0 &=& \frac{2 \Lambda }{1+ \Lambda }
\frac{ \prod_{l=1}^n (1+C_{2l-1}) }{ \prod_{l=1}^{n-1} ( 1+C_{2l} ) } \\
B_l &=& D_0 \frac{ \prod_{i=1}^{n-1} ( C_{2i} - C_{2l-1} ) }
                   { \prod_{i=1, i \ne l}^{n} ( C_{2i-1} - C_{2l-1} ) } \ .
\eea

Comparing (\ref{eq:zolo}) with (\ref{eq:zol}), one immediately
sees that it is more advantageous to use the former approximation
than the latter, especially for computing quark propagators,
since only one more matrix multiplication with $ ( h_w^2 + c_{2n} ) $
after the completion of the inner CG loop
would yield about $ 2.5 $ times higher accuracy
than using (\ref{eq:zol}). Although we used (\ref{eq:zol})
for our computations in Ref. \cite{Chiu:2002xm}, we have switched
to (\ref{eq:zolo}) for better accuracy, in our ongoing lattice QCD
computations.

{\footnotesize
\begin{table}
\begin{center}
\begin{tabular}{|c|c|c|c|c|c|c|}
\hline
$ b $
& \multicolumn{6}{c|}{ $ n $ }  \\
\hline
       &   10  &  12   &   14  &  16  &  18  &  20  \\
\hline
\hline
  1000 & $ 5.6 \times 10^{-9}  $ & $ 9.4 \times 10^{-11} $
       & $ 1.6 \times 10^{-12} $ & $ 2.7 \times 10^{-14} $
       & $ 4.6 \times 10^{-16} $ & $ 7.8 \times 10^{-18} $  \\
\hline
  2000 & $ 2.2 \times 10^{-8}  $ & $ 4.8 \times 10^{-10} $
       & $ 1.1 \times 10^{-11} $ & $ 2.4 \times 10^{-13} $
       & $ 5.3 \times 10^{-15} $ & $ 1.2 \times 10^{-16} $  \\
\hline
  3000 & $ 4.5 \times 10^{-8}  $ & $ 1.1 \times 10^{-9} $
       & $ 2.9 \times 10^{-11} $ & $ 7.5 \times 10^{-13} $
       & $ 1.9 \times 10^{-14} $ & $ 5.0 \times 10^{-16} $  \\
\hline
  4000 & $ 7.2 \times 10^{-8}  $ & $ 2.0 \times 10^{-9} $
       & $ 5.7 \times 10^{-11} $ & $ 1.6 \times 10^{-12} $
       & $ 4.6 \times 10^{-14} $ & $ 1.3 \times 10^{-15} $  \\
\hline
  5000 & $ 1.0 \times 10^{-7}  $ & $ 3.1 \times 10^{-9}  $
       & $ 9.4 \times 10^{-11} $ & $ 2.8 \times 10^{-12} $
       & $ 8.6 \times 10^{-14} $ & $ 2.6 \times 10^{-15} $  \\
\hline
  6000 & $ 1.3 \times 10^{-7}  $ & $ 4.3 \times 10^{-9}  $
       & $ 1.4 \times 10^{-10} $ & $ 4.4 \times 10^{-12} $
       & $ 1.4 \times 10^{-13} $ & $ 4.5 \times 10^{-15} $  \\
\hline
  7000 & $ 1.7 \times 10^{-7}  $ & $ 5.7 \times 10^{-9}  $
       & $ 1.9 \times 10^{-10} $ & $ 6.4 \times 10^{-12} $
       & $ 2.1 \times 10^{-13} $ & $ 7.2 \times 10^{-15} $  \\
\hline
  8000 & $ 2.1 \times 10^{-7}  $ & $ 7.1 \times 10^{-9}  $
       & $ 2.5 \times 10^{-10} $ & $ 8.7 \times 10^{-12} $
       & $ 3.0 \times 10^{-13} $ & $ 1.1 \times 10^{-14} $  \\
\hline
  9000 & $ 2.4 \times 10^{-7}  $ & $ 8.7 \times 10^{-9}  $
       & $ 3.1 \times 10^{-10} $ & $ 1.1 \times 10^{-11} $
       & $ 4.1 \times 10^{-13} $ & $ 1.5 \times 10^{-14} $  \\
\hline
 10000 & $ 2.8 \times 10^{-7}  $ & $ 1.0 \times 10^{-8}  $
       & $ 3.9 \times 10^{-10} $ & $ 1.4 \times 10^{-11} $
       & $ 5.3 \times 10^{-13} $ & $ 2.0 \times 10^{-14} $  \\
\hline
\end{tabular}
\end{center}
\caption{
The deviation
$ d(r^{(n-1,n)}_Z) = \max | 1 - \sqrt{x} r^{(n-1,n)}_Z(x) |
= ( 1-\Lambda )/(1+\Lambda ) $
of the Zolotarev optimal rational approximation (\ref{eq:rZ'})
versus the upper bound $ b $ of $ x \in [ 1, b ] $,
and the degree $ n $ of the rational polynomial.}
\label{table:drZ'}
\end{table}
}

\section{Concluding remarks}

In this paper, we have discussed
the basic principles underlying the rational approximation,
and shown explicitly that the Zolotarev approximation is indeed
the optimal rational approximation for the inverse square root function.
For the overlap Dirac operator, we have derived
the theoretical error bound for the matrix-vector multiplication
$ H_w ( H_w^2 )^{-1/2}_Z Y $, which is equal to two times of the
maximum deviation $ d_Z(n,b) $ of the Zolotarev rational polynomial.
This is also the upper bound for the chiral symmetry breaking due to
Zolotarev approximation.
Some numerical values of $ d_Z(n,b) $ are listed in Table \ref{table:drZ}
as well as plotted in Figure \ref{fig:dev_zolo}.
An empirical formula for $ d_Z(n,b) $ is determined, which provides a
reliable estimate of the theoretical error bound, especially for the
range of parameters : $ b \simeq 10^3 - 10^4 $ and $ n \simeq 10 - 20 $.
We also compare two possible forms of Zolotarev optimal rational
approximation : $ r_Z^{(n,n)}(x) $
and $ r^{(n-1,n)}_Z(x) $, and point out that the former seems to be
the better choice for computing quark propagators, since with the
same computational cost, one has
$ d(r^{(n,n)}_Z) \simeq 0.4 \times d(r^{(n-1,n)}_Z) $.

With Zolotarev optimal rational approximation for $ (H_w^2)^{-1/2} $
in the overlap Dirac operator, one has no difficulties to preserve
exact chiral symmetry to very high precision
( e.g., $ \Delta_Z < 10^{-12} $ ), for any gauge configurations
on a finite lattice. This feature is vital for lattice QCD to
extract physical observables from the first principles.
In practice, one might have difficulties to push the error
$ \sigma $ (\ref{eq:sigma}) down below $ 10^{-13} $, which is essentially
due to the inaccuracies of the projected ( high and low-lying )
eigenmodes of $ H_w^2 $,
rather than the Zolotarev approximation of $ (H_w^2)^{-1/2} $.
In the future, we will try to improve the accuracy of the projected eigenmodes.
In the meantime, the precision of exact chiral symmetry up to $ 10^{-13} $
should be sufficient for many calculations in lattice QCD with overlap Dirac
quarks.




This work was supported in part by the National Science Council,
ROC, under the grant number NSC90-2112-M002-021,
and also in part by NCTS.

\bigskip
\bigskip

\vfill\eject


\end{document}